# Quantitative Chemical Exchange Saturation Transfer Imaging with Golden-Angle Radial k-Space and Locally Low-Rank Reconstruction

Ouri Cohen*[1], Elizabeth J. Sutton[2], Robert J. Young[2], Ricardo Otazo[1,2]

[1]Department of Medical Physics, Memorial Sloan Kettering Cancer Center, New York, NY, USA

[2]Department of Radiology, Memorial Sloan Kettering Cancer Center, New York, NY, USA

***Correspondence to:** Ouri Cohen, coheno1@mskcc.org, Memorial Sloan Kettering Cancer Center, 320 East 61st St, New York, NY, 10025, USA.

**Word Count: 7284**

# Abstract

Chemical exchange saturation transfer magnetic resonance fingerprinting (CEST-MRF) is a promising quantitative molecular imaging technique. To reduce scan time, current CEST-MRF implementations typically use echo-planar imaging (EPI) readouts, which are prone to geometric distortion and signal dropout. In this study, we developed a motion-robust, geometrically accurate CEST-MRF method using radial k-space sampling, locally low-rank reconstruction, and neural network–based quantification. The acquisition schedule was optimized using deep learning, and its accuracy was validated in numerical simulations with digital phantoms. The number of spokes per measurement was determined through simulations and in vivo ablation studies in healthy volunteers. Five healthy subjects underwent repeated scans, and regions of interest were defined for analysis. Tissue maps generated with the proposed method were compared with values obtained from nonlinear least-squares (NLS) fitting of multi-saturation-power z-spectra. Motion sensitivity and test–retest reproducibility were evaluated using the coefficient of variation (CV) and intraclass correlation coefficient (ICC). Clinical feasibility was demonstrated in a subject with a history of remote middle cerebral artery infarction. The results show that 3D quantitative CEST maps can be acquired in 11 minutes using 34 spokes per measurement. Numerical simulations yielded mean errors below 14% for all tissue parameters, while in vivo parameter estimates agreed well with both NLS-derived values and prior brain CEST-MRF studies. The mean ICC across all tissue maps was 0.92 in white matter and 0.87 in gray matter, with mean inter-subject CVs of 5.4% and 3.4%, respectively. For the radial acquisition, the mean error between motion and no-motion conditions was 8.6%. In the pathology case, changes in CEST-MRF tissue parameters were consistent with clinically diagnosed cystic gliosis. This approach enables accurate and reproducible 3D quantitative CEST imaging of the brain within clinically feasible scan times.

# 1. Introduction

Chemical exchange saturation transfer (CEST) MRI is a molecular imaging technique that indirectly detects labile solute protons via their influence on the bulk water signal [1]. Due to the high concentration of the bulk water pool, CEST enables the detection of solute proton concentrations in the millimolar range at high spatial resolutions (~1mm) without the prolonged scan times required in MR spectroscopy. The CEST contrast is generated by applying saturation pulses at different resonance offsets, yielding a spectrum of signal intensities known as the “Z-spectrum” that reflects the contributions from distinct proton moieties (e.g. hydroxyl, amine, amide). Amide protons are particularly relevant in pathologies, as their exchange rate is sensitive to pH changes and can serve as a valuable biomarker [2], [3]. However, conventional CEST acquisitions are limited by long scan times, complex data processing and semi-quantitative contrast measures such as the magnetization-transfer ratio asymmetry [4].

To address these limitations, quantitative CEST methods have recently emerged, leveraging the MR fingerprinting (MRF) framework [5] combined with deep learning optimization and quantification [6], [7], [8], [9], [10], [11], [12]. This integration enables accurate, reproducible whole-brain quantitative CEST mapping in ~6 minutes using an EPI readout for k-space sampling [13]. While EPI offers high signal-to-noise efficiency and unmatched acquisition speed, it also suffers from susceptibility artifacts [14] which arise at air-tissue interfaces such as the temporal lobes and the orbitofrontal cortex in the brain or the boundary between the anterior chest wall and breast tissue. The use of EPI is also restricted for patients with post-surgical metallic staples or ventricular catheters where large susceptibility gradients can cause significant signal dropout (**Figure 1**). Moreover, multi-slice imaging in the body region can be challenging due to through-plane motion.

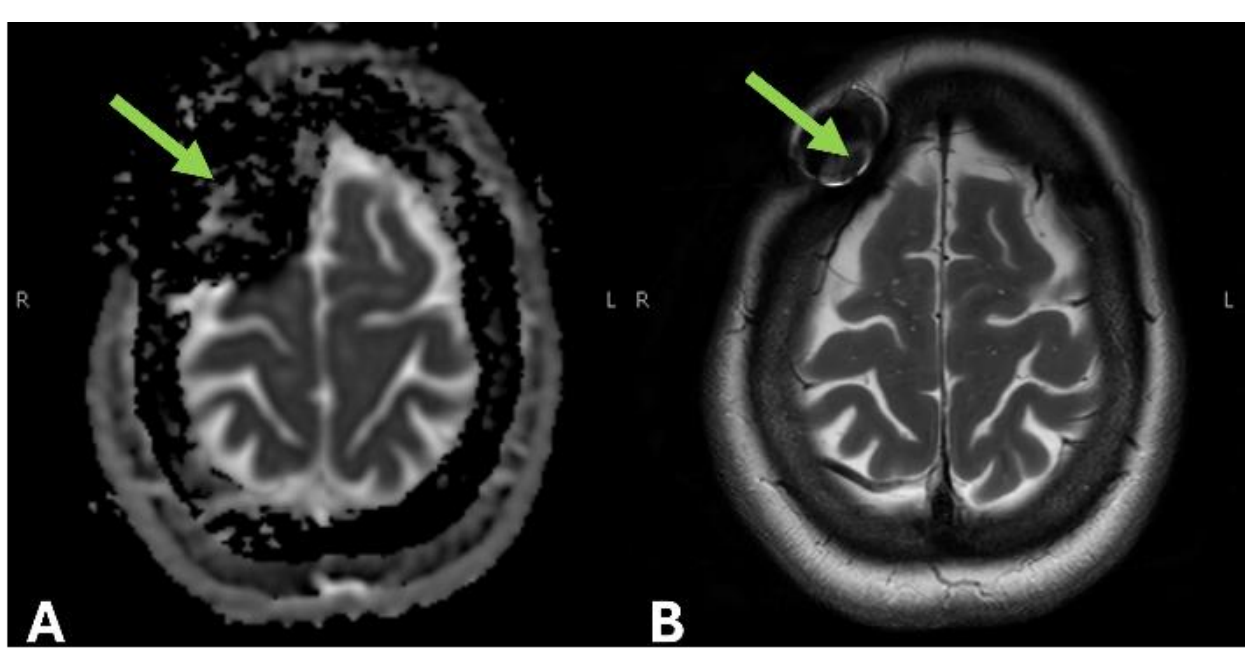

***Figure 1****: Illustration of the impact of a ventricular catheter on an (A) EPI and (B) non-EPI sequence. Green arrows: the presence of the catheter-induced large artifacts and signal dropouts in the EPI image that were not localized to the catheter location.*

Radial k-space sampling presents a promising alternative to overcome EPI-related challenges. First, radial k-space lines enable a very short TR and TE (in the order of a few ms), which minimizes B0 inhomogeneity effects. Second, by oversampling the k-space center and acquiring data along radial trajectories, radial sampling is inherently robust to motion and well-suited for dynamic processes [15]. Furthermore, golden-angle radial trajectories [16] enable efficient k-space coverage, allowing signal recovery using sparse data acquisitions (i.e. reduced scan times) using compressive-sensing reconstruction approaches [17]. Although radial sampling has been applied to conventional MRF sequences [18], adapting it for quantitative CEST imaging requires careful optimization of pulse sequences and acquisition schedules to maintain accuracy while minimizing scan duration.

This work develops an accurate and quantitative radial CEST pulse sequence and reconstruction framework for 3D brain imaging. In a proof-of-concept study, we present a schedule-optimized radial CEST protocol that generates quantitative 3D CEST maps in approximately 11 minutes. We evaluate the accuracy and precision of the proposed method using digital phantoms and in vivo imaging in healthy subjects, and compare its performance with existing quantitative CEST approaches. Finally, we demonstrate its potential clinical utility in a subject with pathology.

# 2. Methods

## 2.1. Pulse-sequence overview

The proposed pulse sequence is illustrated in Figure 2A for one time point in the acquisition schedule. The sequence consists of a variable train of 16ms long, Gaussian-shaped, RF saturation pulses with saturation power $B1_{sat}$ and total saturation time $T_{sat}$. Following saturation, S=30 radial spokes, one for each slice, are sequentially excited using the same excitation flip angle (FA) and spoke angle ϕ. The excited magnetization is allowed to evolve for the duration of the repetition time (TR) before proceeding to the next measurement. After N=30 measurements, the acquisition is repeated P times with the same schedule but with a dynamically incremented spoke angle. The series of spoke angles is defined as: $\varphi_{kp} = G \bullet (P \bullet k + p)$ where $\phi_{kp}$ is the spoke angle in measurement $k$ and in repetition $p$ and G=111.25° is the golden-angle [19]. This ensures a uniform coverage of k-space in each measurement for any given number of spokes by assigning a sequential set of P golden-angle spokes to each measurement (Figure 2B). Specifically, spokes from all P repetitions of each measurement were grouped together to form each of the N images obtained.

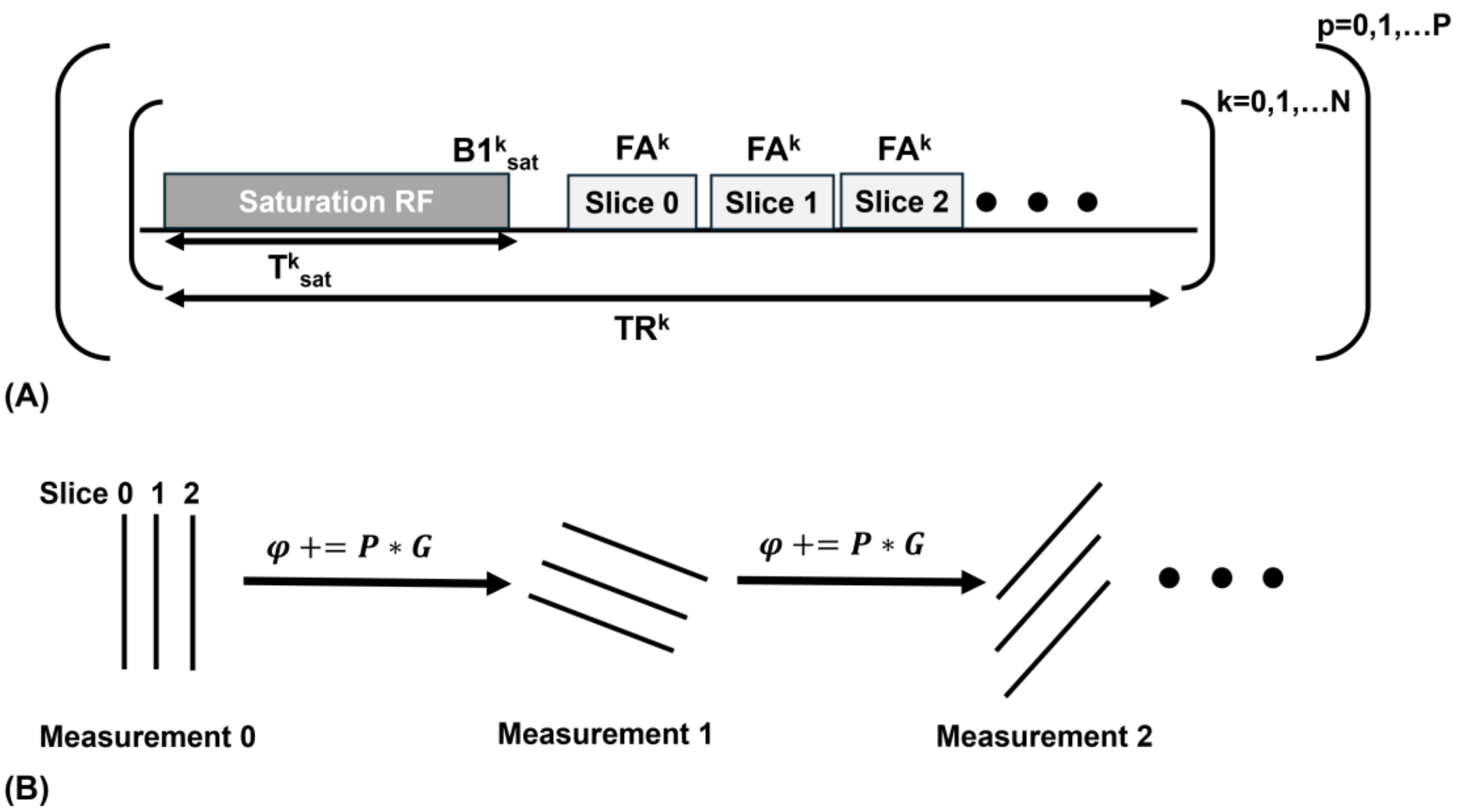

***Figure 2***: *(A) Schematic of the proposed radial CEST-MRF pulse sequence. In each k-th measurement, multiple slices are excited following saturation, with magnetization evolving for time TR before the next measurement. The acquisition is repeated P times with updated spoke angle ϕ following all measurements. (B) To ensure consecutive spokes in each measurement, the spoke angle is updated by integer multiples of the golden-angle G.*

## 2.2. Image reconstruction

Acquired data was reconstructed slice-by-slice. Coil compression was applied to reduce the amount of k-space data from the $C_R$ receiver channels ($C_R$=44) into 8 eigenmodes based on principal component analysis (PCA) [20]. Coil sensitivity map were estimated using the adaptive array-combination technique with coil-reference data given by the temporal average of all acquired spokes [21]. Acquired data were sorted into temporal frames by grouping spokes separated by consecutive multiples of the golden-angle (111.25°). Image reconstruction used a temporal locally-low-rank (LLR) constraint by enforcing each image block over time to have a low-rank structure, as in dynamic imaging [22], and was formulated as the following optimization problem:

$$\min_{m} || E \bullet m - d||_2^2 + \lambda \sum_{i} ||R_i \bullet m||_*$$

where *m* is the image series to be reconstructed, *d* is the measured k-space data, $E=F\cdot C$ is the imaging operator which includes the non-uniform FFT operator *F* [23] as well as the coil sensitivities *C*, $R_i$ selects an image block around pixel *i* for each time and reshapes the block into a local space-time matrix, λ is the regularization weight that controls the tradeoff between parallel imaging data consistency and the low-rank constraint, and $\|M\|_*$ is the nuclear-norm operator to enforce low-rank on the matrix M. The value of λ was empirically set to 1% of the maximum of the signal. While small patches provide lower rank vectors, they also increase the reconstruction time. To limit the reconstruction time, the patch size was therefore set to 128×128 with strides of 64 in each dimension. Equation 1 was solved using nonlinear conjugate gradient descent with a smoothed L1 norm [24] using the approximations $|x| \approx \sqrt{x^2 + \epsilon}$ and $\frac{d|x|}{dx} \approx \frac{x}{\sqrt{x^2+\epsilon}}$ since the L1 norm of the singular values is equivalent to the nuclear norm [25]. A 1D ramp filter (density compensation factor) was applied to each spoke to compensate for variable density sampling.

## 2.3. Schedule optimization

The dual-network method described in Ref. [7] was used to find an optimal scan parameter schedule. In brief, the method includes (i) sampling of the acquisition and tissue parameters, (ii) error calculation for each training schedule, (iii) surrogate network training, and (iv) schedule optimization. The scan and tissue parameters used to train the network were sampled from the ranges shown in Table 1 using Latin hypercube sampling [26] for even coverage of the scan parameter space. The $T_{sat}$ and TR limits were defined to limit the total scan time given the number of repetitions required with radial sampling. The tissue parameters sampled included the water T1 and T2 relaxation (T1w, T2w), the amide exchange rate and volume fraction (ksw, fs), the semisolid exchange rate and volume fraction (kssw, fss), and the transmit field inhomogeneity (B1), which was computed as a fraction of the nominal FA. A set of 2000 acquisition schedules was selected, each consisting of 30 time points. The training dataset used was defined by randomly sampling 10 000 entries from the tissue-parameter ranges given in Table 1. A set of 400 candidate schedules were generated by random sampling from the acquisition parameters (Table 1) and used as the initialization points for a pattern-search optimizer [27] to avoid local minima in the optimization. The optimizer's objective function cost was equally weighted for all acquisition parameters and points exceeding the ranges shown in Table 1 were assigned an infinite cost. The schedule with minimum cost was used for subsequent experiments. Schedules were optimized using a cluster of 8 RTX A5500 GPUs (Nvidia, Santa Clara, CA, USA) with 24 GB of memory.

**Table 1: Acquisition and tissue parameter ranges**

| Acquisition Parameters | Range | Tissue Parameters | Range |
|---|---|---|---|
| Flip Angle (FA) [°] | 30-90 | Water T1 (T1w) [ms] | 1-4000 |
| Repetition Time (TR) [ms] | 200-1000 | Water T2 (T2w) [ms] | 1-3000 |
| Saturation Power ($B1_{sat}$) [μT] | 0-4 | Amide Exchange Rate (ksw) [Hz] | 1-100 |
| Saturation Time ($T_{sat}$) [ms] | 200-750 | Semisolid Exchange Rate (kssw)[Hz] | 1-100 |
| Resonance Offset (ω) [ppm] | 3.5 | Amide Volume Fraction (fs) [%] | 0-0.91 (=0-1000 mM) |
| | | Semisolid Volume Fraction (fss)[%] | 0-13.63 (=0-15 M) |
| | | Transmit Field (B1) [unitless] | 0.4-1.2 |

## 2.4. Numerical simulations

### 2.4.1. Monte-Carlo simulations

Monte-Carlo simulations were performed to assess the quality of the optimized radial CEST-MRF acquisition schedule. A digital phantom with 256×256 tissue parameter values was generated by sampling from a Gaussian distribution with realistic physiological means and standard deviations: $\boldsymbol{\mu}_{phantom}$ = [1000 ms, 60 ms, 30 Hz, 25 Hz, 0.5%, 3.63%, 0.9], $\boldsymbol{\sigma}_{phantom}$ = [300 ms, 20 ms, 20 Hz, 10 Hz, 0.18,% 1.8%, 0.24] for the T1w, T2w, ksw, kssw , fs, fss and B1 parameters respectively [28]. Simulated CEST-MRF acquisitions using digital phantom and the optimized schedule were reconstructed using a trained DRONE network [6]. The normalized RMS error (NRMSE) was computed for each tissue parameter as:

$$NRMSE\ (\%) = 100\ \times \frac{\sqrt{\sum_i (E_i - R_i)^2}}{\overline{R}}$$

where $E_i$ and $R_i$ are the estimated and reference values, respectively, and $\overline{R}$ is the mean of the reference values.

### 2.4.2. Tissue map accuracy versus sampling density

The accuracy of the proposed method was evaluated for varying numbers of spokes per measurement (P) using a custom digital phantom created from the McConnell BrainWeb phantom [29]. Radial CEST-MRF acquisitions were simulated for different values of P, and each reconstructed dataset was processed through the image reconstruction pipeline described in section 2.2. The value of the LLR regularization parameter λ was optimized by a logarithmic grid search over the range $\lambda \in [10^{-6}, 10^{2}]$ of the maximum signal for different values of P with the optimal value found used for subsequent reconstructions.

Quantitative tissue parameter maps were generated from these datasets using a DRONE network trained with a training set of 90,000 samples drawn from the tissue parameter ranges in Table 1. The absolute error between reconstructed and reference values was calculated per pixel as Error (%) = 100 × |Reference - Reconstructed| / Reference. A mean error across the phantom was then calculated for each P.

## 2.5. In vivo human studies

All experiments were conducted on a 3T GE Signa Premier (GE Healthcare, Waukesha, WI, USA) with the built-in transmit body coil and 48-channel receiver head coil. Seven healthy volunteers (1 male, 6 female; mean ± SD age: 25.6 ± 1.7 years) and 1 subject with pathology were recruited for this study and provided written informed consent in accordance with the local Institutional Review Board and Privacy Board.

### 2.5.1. MRI

Brain imaging scans were acquired with the following parameters: field of view (FOV)= 280×280 $mm^2$, matrix size = 256 × 256, in-plane resolution = 1.1 × 1.1 $mm^2$, slice thickness = 3 mm, number of slices=30, echo time (TE) =2.65 ms, and bandwidth = 62.5 kHz. Radial data was reconstructed using the LLR pipeline (section 2.2), and tissue maps were quantified with the DRONE network described in section 2.4.1.

### 2.5.2. Selection of spokes per measurement

To determine the optimal number of spokes per measurement (P) and quantify the tradeoff between image quality and scan time, spoke ablation studies were conducted. One healthy subject was scanned using P=64 spokes/measurement. The k-space data were retrospectively undersampled by discarding averages to generate datasets with P=64, 55, 34, 20, 16 and 8 spokes/measurement. Each dataset was independently reconstructed and the corresponding tissue maps quantified. Regions of interest (ROI) in the grey matter (GM) and white matter (WM) were defined and the mean ± SD of the tissue parameters in each ROI were calculated for all six datasets. The value of P that yielded the best compromise between quantitative accuracy, image blurring and total scan time was used for subsequent experiments.

### 2.5.3. Validation against conventional Z-spectrum CEST acquisition

Reconstructed tissue parameter maps from the radial acquisition were used to synthesize simulated CEST spectra, which were compared with experimentally measured Z-spectra in the same subject. The subject was scanned with a conventional CEST sequence with 59 resonance frequency offsets in the -7.3 to 7.3 ppm range. The FA, $B1_{sat}$, TR and $T_{sat}$ were set to 90°, 2 μT, and 8 and 2.56 s, respectively. The total scan length for the conventional CEST acquisition was 8 min. The long TR was chosen to avoid possible T1 contamination between measurements. The CEST-MRF derived tissue parameters obtained with the optimized schedule were used to generate synthetic Z-spectra by simulating a conventional CEST acquisition using the same acquisition parameters used in the actual scan. To facilitate comparison between the synthetic and measured spectra, both were similarly normalized to their respective value far from resonance. ROIs were defined in the WM and GM and the Pearson correlation coefficient and RMSE between the synthetic and measured spectra was calculated for each region.

### 2.5.4. Validation against multi-power Z-spectrum CEST

To further validate the quantitative accuracy of the extracted tissue parameters, an additional experiment was conducted with CEST spectra acquired using variable saturation powers. Specifically, 6 Z-spectra were acquired with saturation powers of 0.5, 1, 1.5, 2, 2.5, and 3 µT. For each power, 59 offsets were acquired ranging from -7.3 to 7.3 ppm. The FA was set to 90° and the TR and $T_{sat}$ set to 6s and 2.56 s. Given the large number of acquisitions, a shorter TR was used in this experiment to limit the total scan time. To correct for the inevitable subject motion over the long acquisition, the images from each offset were co-registered prior to fitting using MATLAB's *imregister()* function (MathWorks, Natick, MA, USA). WM and GM ROIs were defined. The quantitative tissue parameters corresponding to each voxel were obtained by nonlinear least squares (NLS) fitting using the Levenberg-Marquardt algorithm [30] on the concatenated spectra from all six acquisitions. A 4-pool model consisting of Water, CEST, MT, and NOE pools was used with parameters shown in Table S2. To avoid convergence to local minima, a multi-start approach was used with the fitting algorithm initialized at 10 random points for each voxel. The subject was also scanned with the radial CEST-MRF sequence and the tissue parameters values derived from each method were compared. Agreement between the two methods was assessed using Bland–Altman analysis separately for white matter (WM) and grey matter (GM). For each tissue type, the mean value obtained from the two methods was plotted against the difference between the corresponding measurements to evaluate systematic bias and the level of agreement. The mean difference (bias) and the 95% limits of agreement, calculated as the bias ± 1.96 standard deviations of the differences, were determined to quantify the extent of measurement variability between methods.

### 2.5.5. In vivo repeatability

Test–retest experiments were performed in five subjects to assess the in vivo repeatability of the radial CEST-MRF sequence in the healthy volunteers. Each subject underwent two scans with the proposed sequence separated by a 10-min interval during which the subject was removed from the scanner and repositioned. Scan data were reconstructed with the trained DRONE networks. WM and GM segmentation masks were generated by thresholding the T1w map at values of 500-1000 ms for WM and 1000-1500 ms for GM. Tissue-parameter distributions within GM and WM were calculated using these masks. Repeatability was quantified using the coefficient of variation ($CV = 100 \times SD/Mean$) for the test-retest measurements within each subject ($CV_{within\text{-}subject}$) and for the test measurement across all five subjects ($CV_{across\text{-}subjects}$). The intraclass-correlation coefficient (ICC) with 95% confidence intervals was calculated using a two-way mixed-effects model for absolute agreement (ICC(3,1)) [31]. The intraclass-correlation coefficient (ICC) with 95% confidence intervals was calculated using a two-way mixed-effects model for absolute agreement (ICC(3,1)).

#### 2.5.6. Motion sensitivity

Motion sensitivity testing was conducted using two scan conditions: (1) no motion, and (2) controlled head movement for the radial CEST-MRF sequence. The first scan was acquired with the subject instructed to remain still. In the second scan, the subject was instructed to move their head from side to side at regular intervals, rotating ±45° from the initial supine position. The data from each scan were reconstructed and the tissue maps obtained were compared to those from the no-motion scan. The relative error between the motion and no-motion scans was quantified for each tissue parameter map using the expression Error (%) = 100 × | No Motion-Motion| / No Motion. A global error metric was calculated for each map using a trimmed mean excluding the top and bottom 10% of the data to avoid excessive outlier influence.

### 2.6. Subject with pathology

A 31-year-old female with a history of a remote middle cerebral artery infarct was recruited for this study. The subject was scanned using the radial CEST-MRF sequence, and tissue maps were reconstructed as described in Section 2.2. The resulting maps were evaluated in the context of a diagnosis provided by one of the authors (R.J.Y), a trained neuroradiologist, based on conventional morphological imaging.

# Results

## 3. Results

### 3.1. Optimized acquisition schedule

The radial CEST-MRF acquisition schedule produced with the deep learning schedule optimization is shown in Figure 3A. Optimization of the 30-measurement schedule required 2 hours using an 8 GPU cluster, yielding a scan duration of 20 seconds for all measurements for a single spoke/measurement.

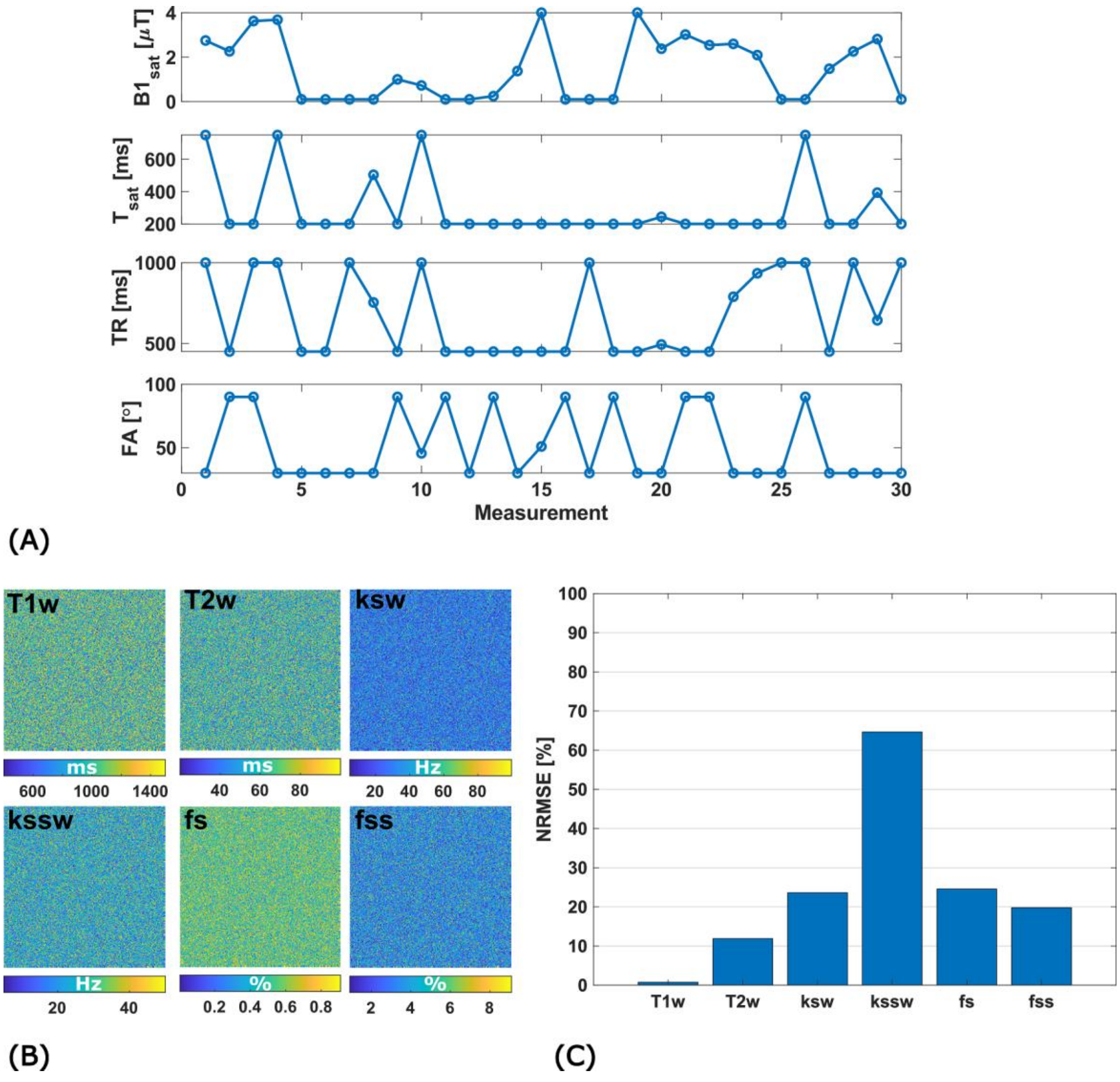

***Figure 3****: (A) Optimized acquisition schedule showing the saturation pulse powers (B1sat), saturation pulse duration (Tsat), the repetition time (TR) and flip angle (FA). (B) Digital tissue phantom for characterization of the optimized schedule. Tissue-parameter values were drawn from a random Gaussian distribution. (C) Normalized RMS error (NRMSE) of reconstructed tissue parameter values from simulations using the optimized schedule.*

## 3.2. Numerical simulations

The tissue parameter values used in Monte Carlo simulations (Figure 3) yielded NRMSE values below 25% for all parameters except kssw. The constant resonance offset used in the acquisition schedule predictably exhibited relatively poor sensitivity to kssw variations. Reconstructed tissue parameter maps for simulated acquisitions with P=256 and P=34 spokes per measurement are compared to true maps in Figure 4A, with corresponding error maps. The optimal value of the LLR regularization parameter λ for all spokes per measurements is shown in Figure S1. The optimal value of λ decreased with increasing number of spokes per measurement, as expected. The mean error across the digital phantom for different P values is plotted in Figure 3B, showing that P=34 achieved <13% error for all parameters.

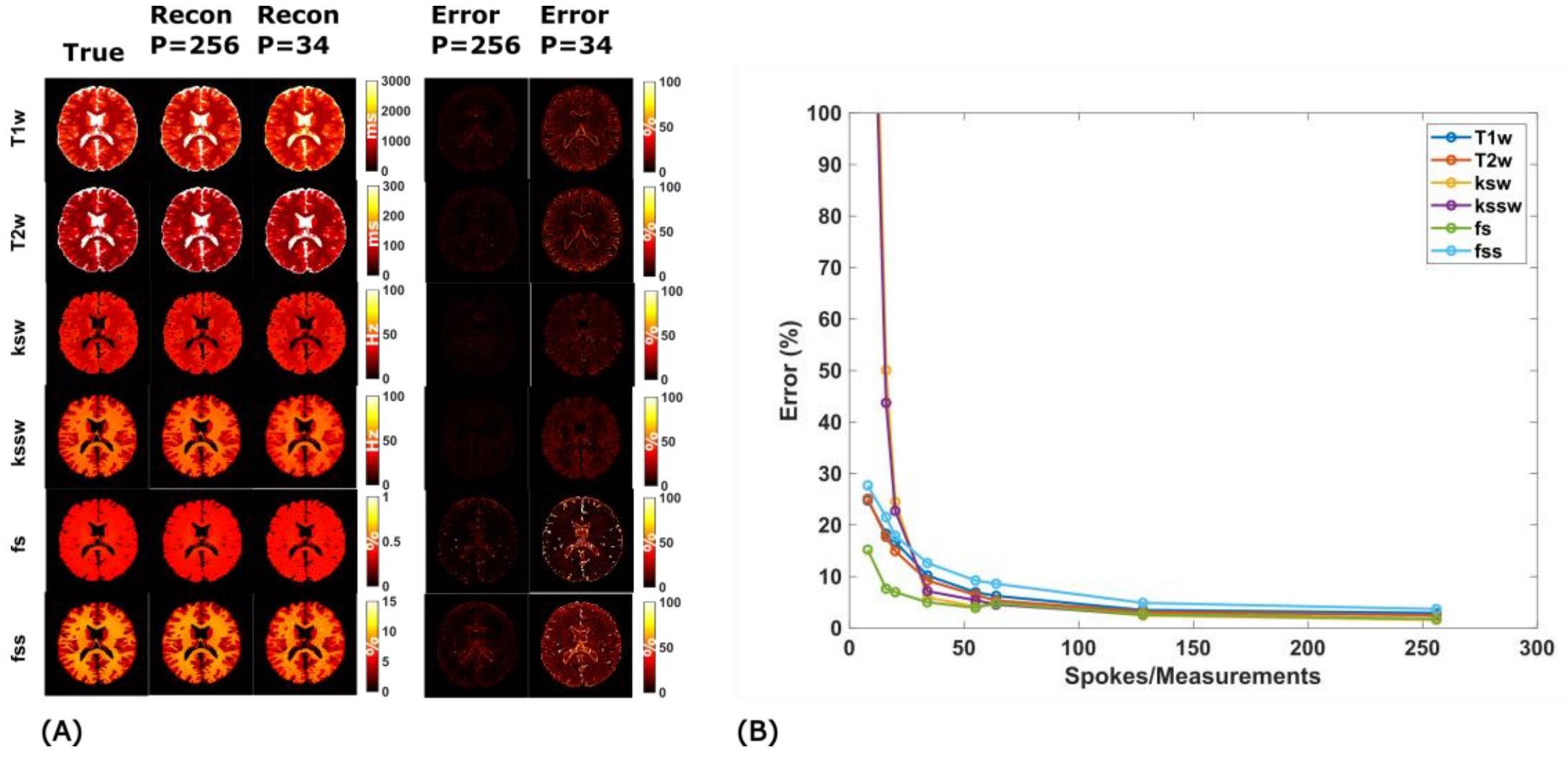

***Figure 4***: *(A) Tissue parameter maps from simulations using P=256 and P=34 spokes per measurements in a digital brain phantom. (B) Mean error across tissue parameters for different spokes per measurements.*

## 3.3. Human brain studies

### 3.3.1. In vivo spoke ablation

Tissue parameter maps from retrospectively undersampled acquisitions are illustrated for a representative slice in Figure S2. Mean ±SD values for WM and GM regions across all undersampled acquisitions are shown in Figure S3. Reducing spokes per measurement from P=64 to P=55 resulted in negligible changes to tissue maps or parameter values, whereas reducing to P=8 caused severe blurring and inaccurate parameter values. Therefore, P=34 was selected for subsequent experiments as it minimized blurring and parameter errors while maintaining an acceptable scan time of 11 minutes.

### 3.3.2. Validation against conventional CEST

The synthesized CEST curves from both scans for the GM and WM regions are shown overlaid on the measured CEST curves in Figure 5. There was overall good agreement between the measured and synthesized curves with a Pearson correlation of 0.97 for WM and 0.99 for GM and RMSE of 0.0401/0.0276 for WM/GM. Differences in the curves in the negative offset regions were due to NOE effects that were not included in the CEST-MRF model leading to the differences between the curves in the negative offset region.

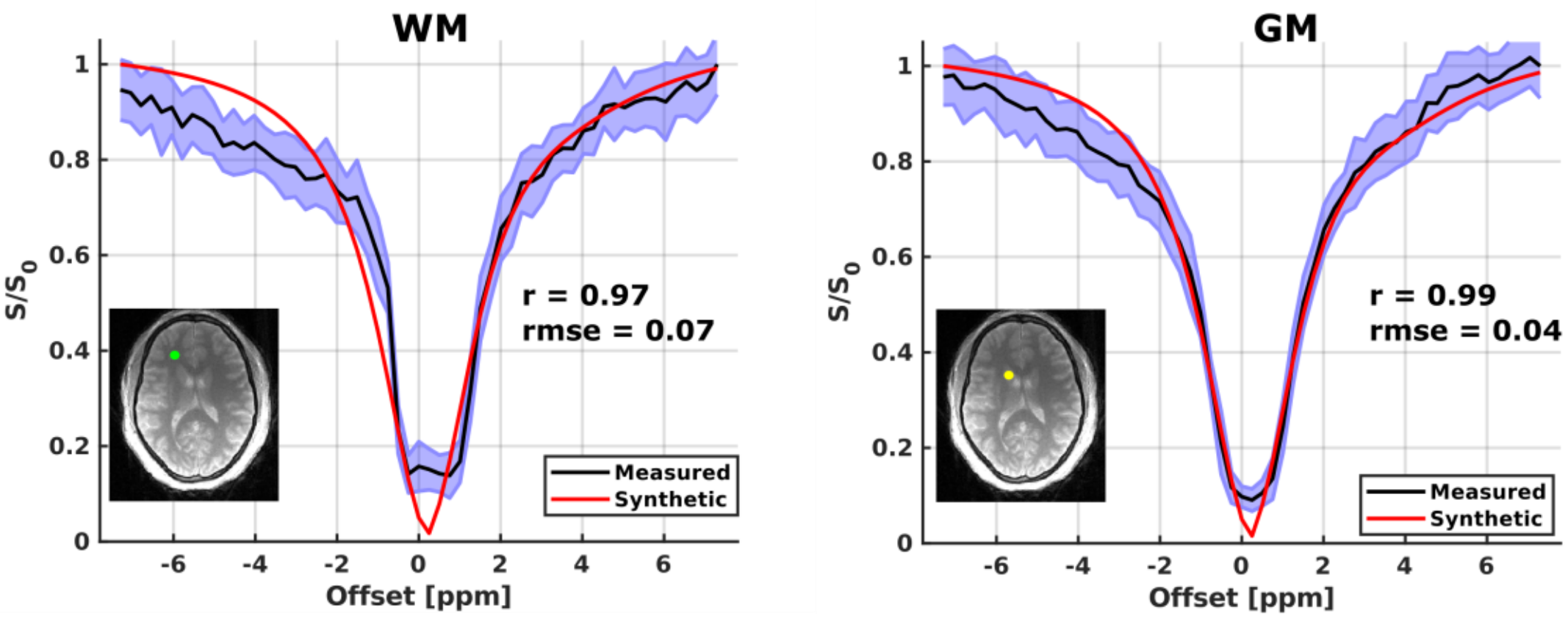

***Figure 5****: Comparison between a measured CEST spectrum (black line) and one synthesized from the CEST-MRF parameters (red line) obtained with the optimized schedule for WM and GM. The location of the ROI is shown inset for WM (green region) and GM (yellow region). The blue shading indicates the standard deviation of the spectra in the given ROI. The measured and synthetic curves were highly correlated (r > 0.97). NOE effects were present in the measured data but not included in the CEST-MRF model leading to the differences between the curves in the negative offset region.*

### 3.3.3. Validation against multi-power Z-spectrum CEST

The measured and fitted curves for the multi-power Z-spectra are shown in Figure 6 for the WM and GM ROIs shown inset. The tissue parameter values obtained via NLS fitting are listed in Table 3 in comparison to those from the proposed radial CEST-MRF sequence in a representative subject. The Bland-Altman plots for all tissue parameters in are shown for WM in Figure S4 and for GM in Figure S5. Overall, good agreement was observed between the two methods for all extracted parameters. However, the CEST-MRF derived tissue parameter values exhibited a reduced variance compared to the multi-power fitting approach, which also required a significantly longer scan time (~35 min vs 11 min) and processing time (~2 hours for fitting each ROI on a 12-core CPU).

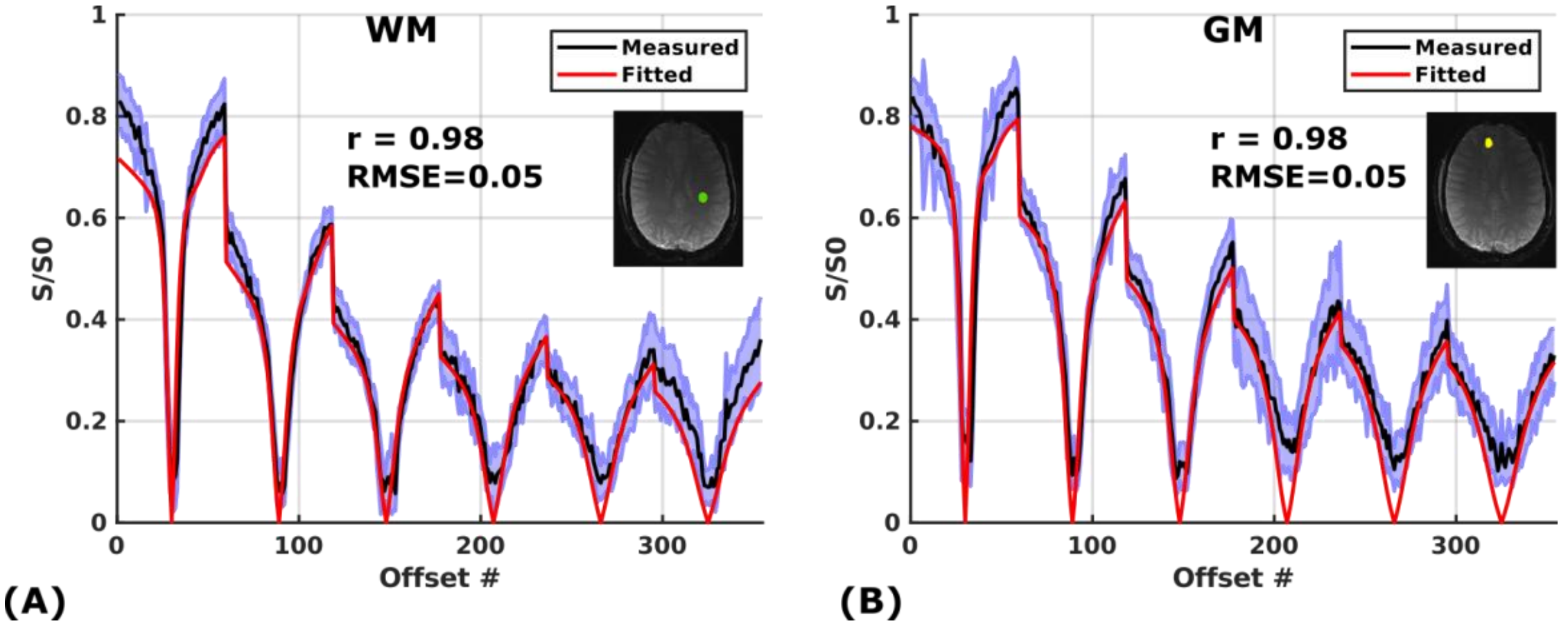

***Figure 6**: Comparison between the measured multi-power CEST spectrum (black line) and one obtained from NLS fitting to a 4-pool model (red line) for WM and GM. The location of the ROI is shown inset for WM (green region) and GM (yellow region). The blue shading indicates the standard deviation of the spectra in the given ROI. The measured and fitted curves were highly correlated (r = 0.98).*

### 3.3.4. Volumetric tissue maps in healthy subjects

Volumetric tissue parameter maps for a representative subject are shown in Figure 7. Axial, coronal and sagittal cross-sections were generated by interpolating the measured 256×256×30 voxel data into a $256^3$ matrix using MATLAB's *imresize3()* function. Mean ± SD values for GM and WM tissue parameters across the five subjects, along with NLS fitting values from the multi-power Z-spectrum CEST validation, are listed in Table 2.

**Table 2**: Mean ± SD tissue parameter values and repeatability metrics for WM and GM ROIs using radial CEST-MRF.

| Parameter | WM Mean ± SD[a] | WM $CV_{within-subject}$ [%] | WM $CV_{across-subjects}$ [%] | WM ICC [95% CI] | GM Mean ± SD[a] | GM $CV_{within-subject}$ [%] | GM $CV_{across-subjects}$ [%] | GM ICC [95% CI] |
|---|---|---|---|---|---|---|---|---|
| T1w [ms] | 775.88 ± 19.37 | 0.16 | 2.50 | 0.99 [0.94, 1] | 1200.1 ± 7.43 | 0.06 | 0.62 | 0.99 [0.87, 1] |
| T2w [ms] | 68.59 ± 3.95 | 1.22 | 5.75 | 0.90 [0.34, 0.99] | 86.67 ± 3.75 | 1.04 | 4.33 | 0.9 [0.31, 0.99] |
| ksw [Hz] | 33.31 ± 1.45 | 1.39 | 4.36 | 0.90 [0.34, 0.99] | 37.90 ± 0.94 | 0.86 | 2.49 | 0.86 [0.18, 0.98] |
| kssw [Hz] | 39.69 ± 2.62 | 1.67 | 6.60 | 0.87 [0.21, 0.99] | 35.70 ± 1.61 | 1.58 | 4.52 | 0.67 [0, 0.96] |
| fs [%] | 0.40 ± 0.01 | 0.77 | 3.51 | 0.91 [0.37, 0.99] | 0.37 ± 0.008 | 0.52 | 2.23 | 0.93 [0.48, 0.99] |
| fss [%] | 6.51 ± 0.63 | 1.12 | 9.70 | 0.97 [0.75, 1] | 5.31 ± 0.33 | 1.70 | 6.19 | 0.85 [0.14, 0.98] |

[a]calculated over n=5 healthy subjects

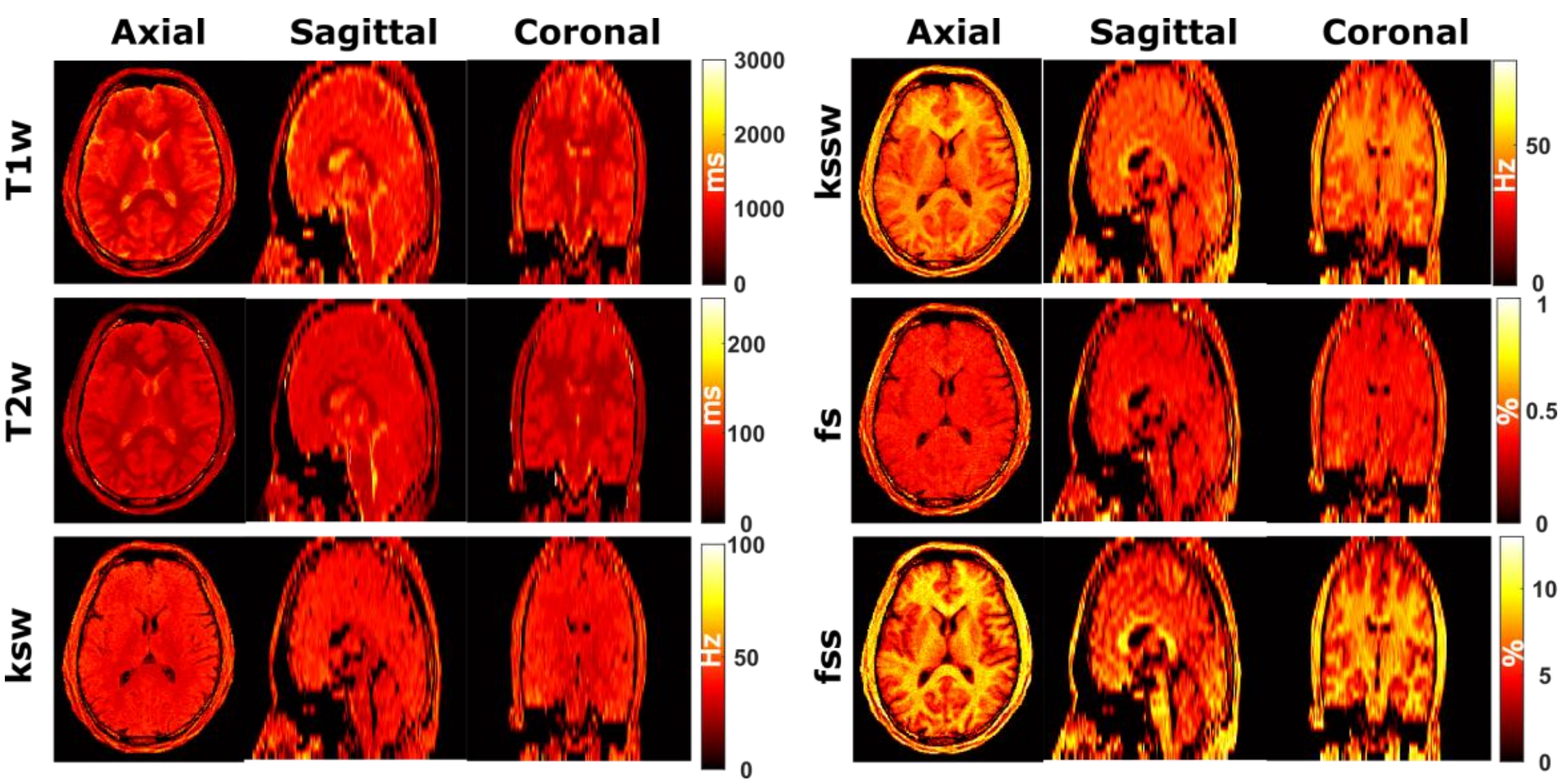

***Figure 7****: Axial, sagittal and coronal cross-sections from a whole-brain radial acquisition with optimized schedule. fs, volume fraction; fss, volume fraction; kssw, semisolid exchange rate; ksw, amide exchange rate; T1w, water T1; T2w, water T2.*

**Table 3**: Comparison of tissue parameter values from NLS fitting and radial CEST-MRF for WM and GM.

| Parameter | WM NLS | WM CEST-MRF | GM NLS | GM CEST-MRF |
|---|---|---|---|---|
| T1w [ms] | 932.2 ± 164.2 | 870.32 ± 89.92 | 1180 ± 49.23 | 1111.9 ± 69.85 |
| T2w [ms] | 78.41 ± 16.49 | 68.73 ± 3.80 | 81.34 ± 6.455 | 78.70 ± 4.64 |
| ksw [Hz] | 49.17 ± 34.02 | 34.81 ± 4.48 | 42.69 ± 19.17 | 39.86 ± 2.71 |
| kssw [Hz] | 46.58 ± 13.64 | 40.50 ± 2.90 | 34.73 ± 1.272 | 38.80 ± 2.27 |
| fs [%] | 0.42 ± 0.03 | 0.40 ± 0.05 | 0.43 ± 0.06 | 0.37 ± 0.03 |
| fss [%] | 5.92 ± 0.02 | 7.14 ± 0.63 | 4.58 ± 0.12 | 5.72 ± 0.56 |

### 3.3.5. In vivo repeatability

The tissue maps from the test-retest experiment are shown in Figure 8A for a sample central slice. The distribution of tissue parameter values in the GM and WM regions are shown in Figure 8B. The CV and ICC values obtained in each ROI are listed in Table 2. For simplicity, the $CV_{within\text{-}subject}$ in Table 2 is shown as a mean over all subjects. The average over all tissue parameters of the $CV_{within\text{-}subject}$ was 1.05% for WM and 0.96% for GM. The average of the $CV_{across\text{-}subject}$ across all tissue parameters was 5.4%/3.4% for WM/GM. The average ICC across all tissue parameters was 0.92% for WM and 0.87 for GM.

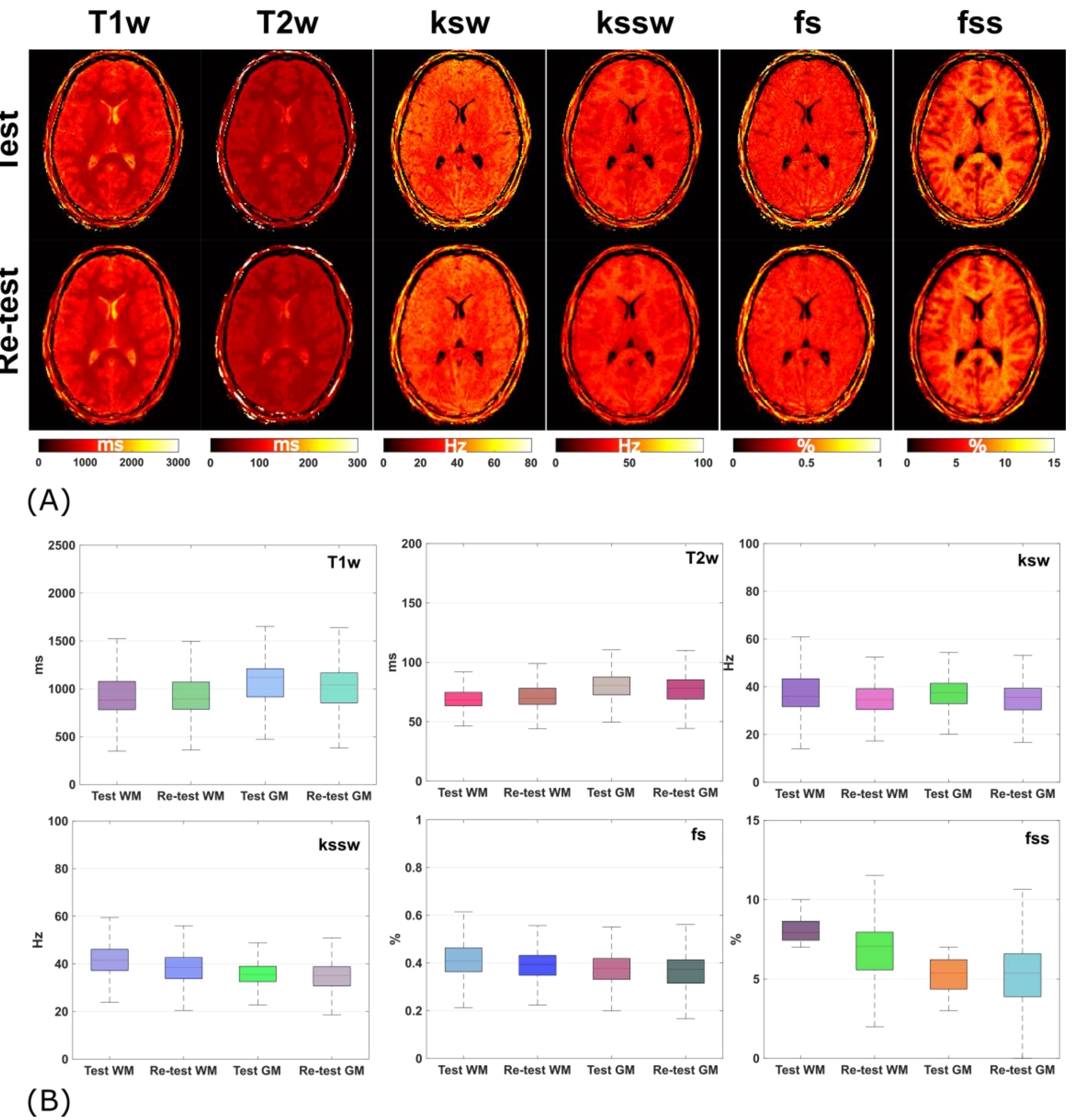

***Figure 8****: Comparison between tissue parameter values obtained in the test-retest experiment. (A) Quantitative tissue maps obtained in each scan in a sample slice. Note the similarity between the maps. (B) Box and whiskers plot of the tissue parameter values for the whole-brain GM and WM ROIs. The two acquisitions showed good agreement between the two scans taken 10 minutes apart.*

## 3.4. Motion sensitivity

Tissue maps for the scan with and without motion using the radial sequence are shown in Figure 9 along with the corresponding error maps. The mean error in the radial maps was 8.48, 6.87, 8.92, 6.87, 9.30 and 11.38% for T1w, T2w, ksw, kssw, fs and fss respectively, and 8.6% averaged over all parameters.

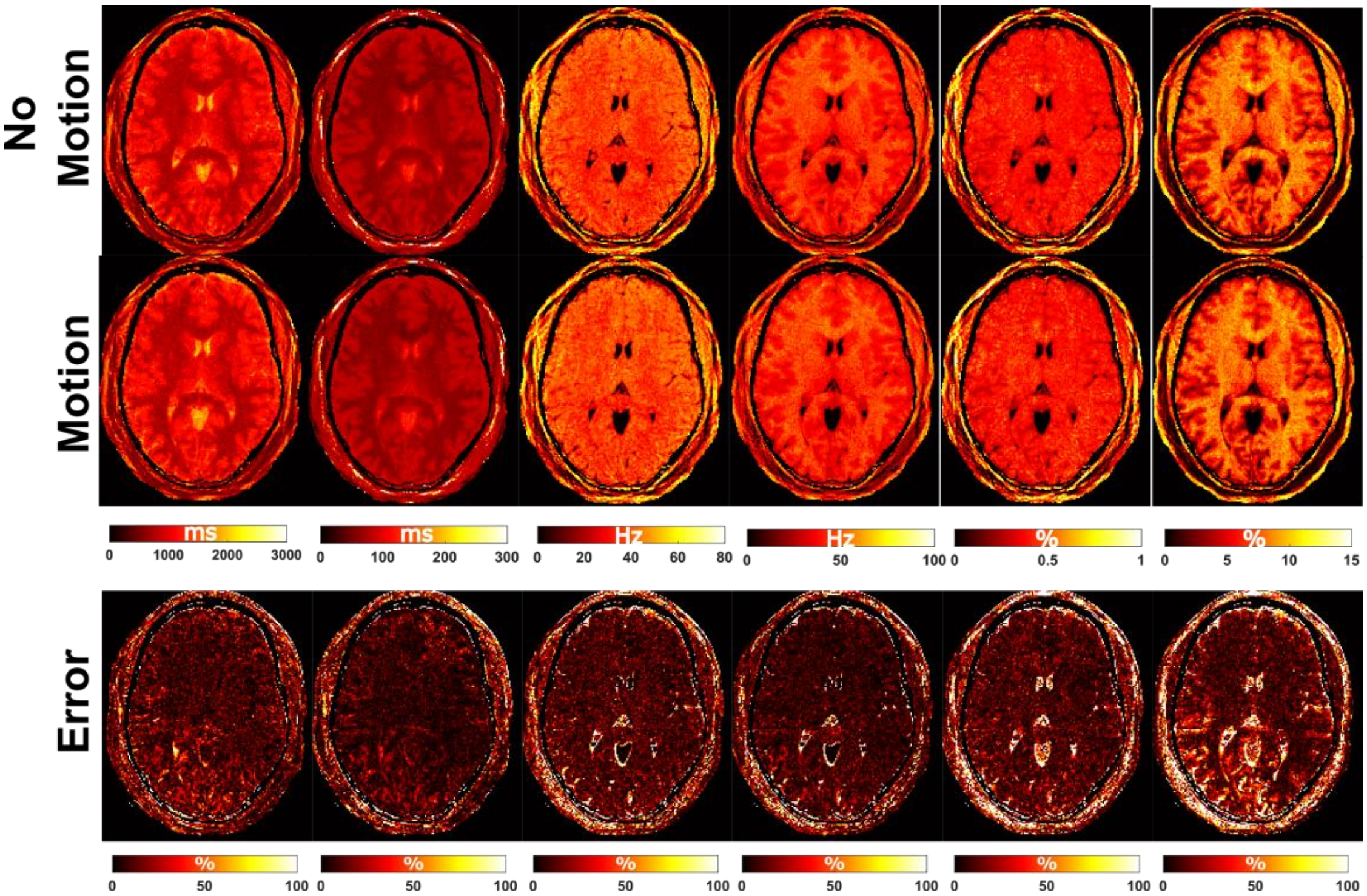

***Figure 9**: Comparison of the motion sensitivity. Tissue and error maps obtained with the radial sequence with and without motion. Note the small error between the no-motion and motion scans provided by the radial sequence.*

## 3.5. Subject with pathology

Tissue parameter maps obtained with the radial CEST-MRF sequence in the pathology subject are shown in Figure 10A and corresponding clinical images obtained with an MP-RAGE sequence shown in Figure 10B.

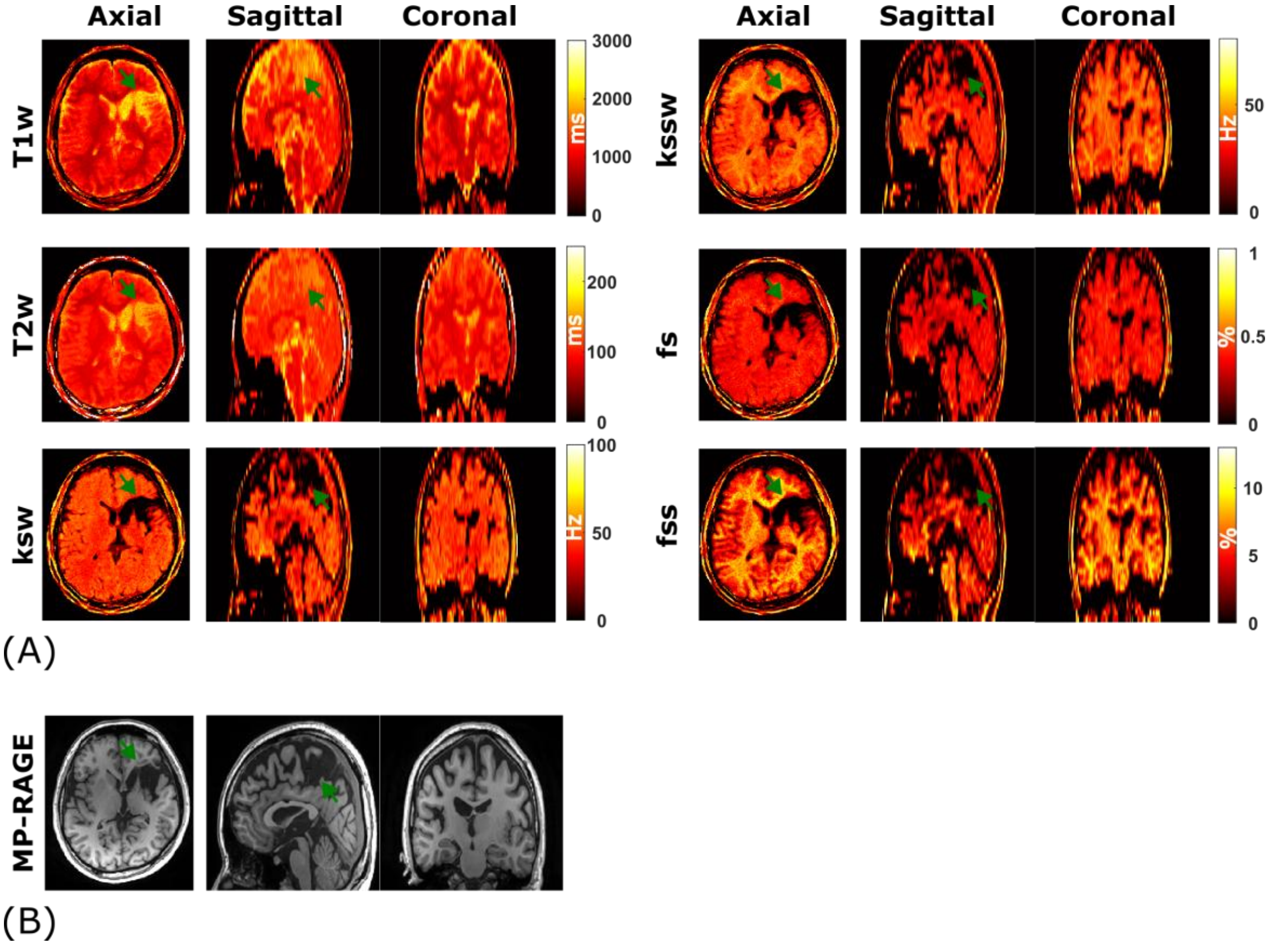

***Figure 10****: Axial, sagittal and coronal cross-sections in a subject with an old infarct. (A) tissue parameter maps obtained with the radial sequence. The large water relaxation values (T1w, T2w) and small CEST parameter values (ksw, kssw, fs,fss) are consistent with CSF filled areas of gliosis. (B) Corresponding images obtained with a clinical MP-RAGE sequence.*

# Discussion

## 4. Discussion

This work demonstrated a proof-of-concept for a motion-robust and spatially accurate CEST-MRF acquisition method using radial k-space sampling. Radial sampling enabled short echo times, successfully mitigating B0-induced geometric distortions and motion artifacts Figure 9). However, the single spoke per TR in radial acquisitions necessitates shorter TRs and saturation pulse durations (Tsat) to minimize scan time which can reduce sensitivity. Monte-Carlo simulations with our optimized schedule yielded an NRMSE (Figure 3) comparable to other CEST-MRF methods [13] and demonstrating high accuracy for most of the tissue parameters, despite the significantly shorter TR and Tsat (Table 1). Although the kssw exhibited larger error, it should be noted that tissue parameter distributions in the Monte Carlo simulations (Figure 3) were drawn from a relatively wide Gaussian distribution. This approach introduces greater variability in parameter values than would typically be encountered in vivo, where biological constraints result in more tightly clustered tissue properties. Consequently, the simulation results likely overestimate the actual errors that would be observed in vivo and the simulations therefore provide a conservative estimate of the method's accuracy. This interpretation is supported by the brain phantom simulations (Figure 4), in which the error in kssw (<10%) was substantially lower than suggested by the Monte Carlo experiments.

We used an intermediate number of spokes per frame to balance scan duration, image blurring, and quantitative accuracy of the tissue parameter maps. In vivo retrospective spoke ablation studies (Figures S2, S3) confirmed that this compromise is acceptable. Although a prospective acquisition with P=34 spokes would use different set of spoke angles than a retrospectively undersampled acquisition with P=34 spokes, this does not affect the reconstruction because golden-angle sampling still provides uniform k-space coverage despite differences in spoke angle distribution. This is further illustrated by the retrospective P=55 reconstruction (Figure S2), which showed little difference from the maps obtained with the prospective P=64 acquisition.

Reconstruction of radially-sampled image time series often uses sparsifying transforms that exploit smooth temporal variations (e.g. contrast enhancement, respiratory motion [32], [33]). However, our optimized schedule produced more complex, discontinuous temporal dynamics, motivating the use of temporal locally low-rank regularization, which improved reconstruction and subsequent quantification. Recent advances in data-adaptive deep-learning reconstructions [34], [35] may further improve image quality, reduce the number of spokes (and thus scan time), or both. These directions will be explored in future work.

Tissue parameter values in healthy subjects were consistent with other CEST-MRF studies using different acquisition schedules and k-space sampling strategies [7,8,11,36,37] as well as other quantitative MRI approaches [38]. The strong agreement between the measured and synthesized Z-spectra in the amide region (Figure 5) provides simultaneous (albeit indirect) validation of the estimated tissue parameters, since all parameters contribute to the synthetized signal. An important distinction of the CEST-MRF approach is its specific saturation of the amide pool, which avoids the confounding effects of NOE contributions that mix into the MTR asymmetry and interfere with conventional CEST quantification. This is a distinct advantage of the CEST-MRF approach over conventional CEST imaging. However, this specificity also means that z-spectra synthesized from CEST-MRF-derived tissue maps will not exhibit the NOE effect, as the method explicitly excludes this contribution, leading to discrepancies between measured and synthesized z-spectra in vivo for that spectral region(Figure 5).

The multi-power spectrum fitting in this work is similar to the QUESP method [39] although with several important differences. First, a four-pool model was used to account for the various proton pools present in vivo. Second, to enable extraction of multiple tissue parameters, data from all resonance offsets was used, rather than the two frequencies used in the original study [39]. Despite the richer dataset, fitting highly nonlinear high-dimensional data with conventional algorithms remains challenging because of the presence of multiple local minima and degenerate solutions. This difficulty is reflected in the relatively large standard deviation observed for several tissue parameters derived using NLS fitting. Although the use of multiple initializations helped mitigate this issue, the required number of initializations was chosen empirically and may not have been optimal. It also substantially increased processing time, making the method impractical for whole-slice fitting.

Nevertheless, the mean radial CEST-MRF derived tissue parameter values were similar to those derived from the Bloch-McConnell fitting of the multi-power Z-spectrum (Table 3). Moreover, Bland-Altman analysis demonstrated good agreement between the two methods in both WM and GM. The mean differences (bias) were small, indicating minimal systematic offset between the methods, and most paired measurements fell within the 95% limits of agreement. Comparable levels of agreement were observed in WM and GM, suggesting that the two methods provide similar parameter estimates across these tissue types.

In addition to improved geometric fidelity, the radial sequence yielded low within-subject and across-subjects coefficients of variations and excellent repeatability (average ICC > 0.87, Table 2) in WM and GM.

This study's relatively small sample size (n=5) is a limitation, and larger cohorts will be required to establish baseline values in healthy subjects. Nevertheless, maps from a patient with pathology (Figure 10) show notable differences from the healthy-subject maps, demonstrating the potential diagnostic utility of the proposed approach. In particular, the clinically diagnosed cystic gliosis in the right insula, subinsula and frontal operculum resulting from the known remote infarct in the middle cerebral artery distribution are consistent with the prolonged T1w and T2w and reduced ksw, kssw, fs and fss parameters in the tissue parameter maps.

# Conclusion

A motion-robust and spatially accurate quantitative radial CEST pulse sequence and reconstruction framework was demonstrated for accurate and reproducible 3D brain quantitative CEST imaging in clinically relevant scan times.

# Acknowledgments

This work was supported by Grant/Award Numbers: P30 CA008748, R37-CA262662. This manuscript is the result of funding in whole or in part by the National Institutes of Health (NIH). It is subject to the NIH Public Access Policy. Through acceptance of this federal funding, NIH has been given a right to make this manuscript publicly available in PubMed Central upon the Official Date of Publication, as defined by NIH.

## Conflict of Interest

A patent application has been submitted on the CEST-MRF technology described in this work.

## References

1. Kyle M Jones, Alyssa C Pollard, Mark D Pagel. Clinical applications of chemical exchange saturation transfer (CEST) MRI. *Journal of Magnetic Resonance Imaging*. 2018;47(1):11-27. doi:10.1002/jmri.25838

2. Dario L Longo, Antonietta Bartoli, Lorena Consolino, Paola Bardini, Francesca Arena, Markus Schwaiger, Silvio Aime. In vivo imaging of tumor metabolism and acidosis by combining PET and MRI-CEST pH imaging. *Cancer research*. 2016;76(22):6463-6470.

3. Annasofia Anemone, Lorena Consolino, Francesca Arena, Martina Capozza, Dario Livio Longo. Imaging tumor acidosis: A survey of the available techniques for mapping in vivo tumor pH. *Cancer and Metastasis Reviews*. 2019;38(1-2):25-49. doi:10.1007/s10555-019-09782-9

4. B Wu, G Warnock, M Zaiss, C Lin, M Chen, Z Zhou, L Mu, D Nanz, R Tuura, G Delso. An overview of CEST MRI for non-MR physicists. *EJNMMI physics*. 2016;3(1):19. doi:10.1186/s40658-016-0155-2

5. Dan Ma, Vikas Gulani, Nicole Seiberlich, Kecheng Liu, Jeffrey L Sunshine, Jeffrey L Duerk, Mark A Griswold. Magnetic resonance fingerprinting. *Nature*. 2013;495(7440):187-192.

6. Ouri Cohen, Bo Zhu, Matthew S Rosen. MR fingerprinting deep reconstruction network (DRONE). *Magnetic resonance in medicine*. 2018;80(3):885-894. doi:10.1002/mrm.27198

7. Ouri Cohen, Ricardo Otazo. Global Deep Learning Optimization of CEST MR Fingerprinting (CEST-MRF) Acquisition Schedule. *NMR in Biomedicine*. 2023;36:e4954. doi:10.1002/nbm.4954

8. Ouri Cohen, Shuning Huang, Michael T McMahon, Matthew S Rosen, Christian T Farrar. Rapid and quantitative chemical exchange saturation transfer (CEST) imaging with magnetic resonance fingerprinting (MRF). *Magnetic resonance in medicine*. 2018;80(6):2449-2463. doi:10.1002/mrm.27221

9. Or Perlman, Kai Herz, Moritz Zaiss, Ouri Cohen, Matthew S Rosen, Christian T Farrar. CEST MR-Fingerprinting: Practical considerations and insights for acquisition schedule design and improved reconstruction. *Magnetic resonance in medicine*. 2020;83(2):462-478.

10. Or Perlman, Bo Zhu, Moritz Zaiss, Matthew S Rosen, Christian T Farrar. An end-to-end AI-based framework for automated discovery of rapid CEST/MT MRI acquisition protocols and molecular parameter quantification (AutoCEST). *Magnetic Resonance in Medicine*. 2022;87(6):2792-2810. doi:10.1002/mrm.29173

11. Or Perlman, Hirotaka Ito, Kai Herz, Naoyuki Shono, Hiroshi Nakashima, Moritz Zaiss, E Antonio Chiocca, Ouri Cohen, Matthew S Rosen, Christian T Farrar. Quantitative imaging of apoptosis following oncolytic virotherapy by magnetic resonance fingerprinting aided by deep learning. *Nature biomedical engineering*. 2022;6(5):648-657. doi:10.1038/s41551-021-00809-7

12. Nikita Vladimirov, Ouri Cohen, Hye-Young Heo, Moritz Zaiss, Christian T Farrar, Or Perlman. Quantitative molecular imaging using deep magnetic resonance fingerprinting. *Nature protocols*. Published online 2025:1-31. doi:10.1038/s41596-025-01152-w

13. Ouri Cohen, Robert J Young, Ricardo Otazo. Quantitative multislice and jointly optimized rapid CEST for in vivo whole-brain imaging. *Magnetic Resonance in Medicine*. 2025;94(2):541-553. doi:10.1002/mrm.30488

14. Peter Jezzard, Robert S Balaban. Correction for geometric distortion in echo planar images from B0 field variations. *Magnetic resonance in medicine*. 1995;34(1):65-73. doi:10.1002/mrm.1910340111

15. Li Feng. 4D Golden-angle radial MRI at subsecond temporal resolution. *NMR in Biomedicine*. 2023;36(2):e4844. doi:10.1002/nbm.4844

16. Li Feng, Robert Grimm, Kai Tobias Block, Hersh Chandarana, Sungheon Kim, Jian Xu, Leon Axel, Daniel K Sodickson, Ricardo Otazo. Golden-angle radial sparse parallel MRI: Combination of compressed sensing, parallel imaging, and golden-angle radial sampling for fast and flexible dynamic volumetric MRI. *Magnetic resonance in medicine*. 2014;72(3):707-717. doi:10.1002/mrm.24980

17. Li Feng. Golden-angle radial MRI: Basics, advances, and applications. *Journal of Magnetic Resonance Imaging*. 2022;56(1):45-62. doi:10.1002/jmri.28187

18. Martijn A Cloos, Jakob Assländer, Batool Abbas, James Fishbaugh, James S Babb, Guido Gerig, Riccardo Lattanzi. Rapid radial T1 and T2 mapping of the hip articular cartilage with magnetic resonance fingerprinting. *Journal of Magnetic Resonance Imaging*. 2019;50(3):810-815.

19. Stefanie Winkelmann, Tobias Schaeffter, Thomas Koehler, Holger Eggers, Olaf Doessel. An optimal radial profile order based on the Golden Ratio for time-resolved MRI. *IEEE transactions on medical imaging*. 2006;26(1):68-76. doi:10.1109/TMI.2006.885337

20. Martin Buehrer, Klaas P Pruessmann, Peter Boesiger, Sebastian Kozerke. Array compression for MRI with large coil arrays. *Magnetic Resonance in Medicine: An Official*

*Journal of the International Society for Magnetic Resonance in Medicine*. 2007;57(6):1131-1139. doi:10.1002/mrm.21237

21. David O Walsh, Arthur F Gmitro, Michael W Marcellin. Adaptive reconstruction of phased array MR imagery. *Magnetic Resonance in Medicine: An Official Journal of the International Society for Magnetic Resonance in Medicine*. 2000;43(5):682-690. doi:10.1002/(SICI)1522-2594(200005)43:5<682::AID-MRM10>3.0.CO;2-G

22. Joshua Trzasko, Armando Manduca, Eric Borisch. Local versus global low-rank promotion in dynamic MRI series reconstruction. In: *Proc. Int. Symp. Magn. Reson. Med*. Vol 19. 2011:4371.

23. Jeffrey A Fessler. On NUFFT-based gridding for non-Cartesian MRI. *Journal of magnetic resonance*. 2007;188(2):191-195. doi:10.1016/j.jmr.2007.06.012

24. Michael Lustig, David Donoho, John M Pauly. Sparse MRI: The application of compressed sensing for rapid MR imaging. *Magnetic Resonance in Medicine: An Official Journal of the International Society for Magnetic Resonance in Medicine*. 2007;58(6):1182-1195. doi:10.1002/mrm.21391

25. Benjamin Recht, Maryam Fazel, Pablo A Parrilo. Guaranteed minimum-rank solutions of linear matrix equations via nuclear norm minimization. *SIAM review*. 2010;52(3):471-501. doi:10.1137/070697835

26. Jeong-Soo Park. Optimal Latin-hypercube designs for computer experiments. *Journal of statistical planning and inference*. 1994;39(1):95-111. doi:10.1016/0378-3758(94)90115-5

27. Charles Audet, John E Dennis Jr. Analysis of generalized pattern searches. *SIAM Journal on optimization*. 2002;13(3):889-903.

28. Andrew Mao, Sebastian Flassbeck, Jakob Assländer. Bias-reduced neural networks for parameter estimation in quantitative MRI. *Magnetic Resonance in Medicine*. Published online 2024.

29. D Louis Collins, Alex P Zijdenbos, Vasken Kollokian, John G Sled, Noor J Kabani, Colin J Holmes, Alan C Evans. Design and construction of a realistic digital brain phantom. *IEEE transactions on medical imaging*. 1998;17(3):463-468. doi:10.1109/42.712135

30. Jorge J Moré. *Levenberg–Marquardt Algorithm: Implementation and Theory*. Argonne National Lab., IL (USA); 1976.

31. Patrick E Shrout, Joseph L Fleiss. Intraclass correlations: Uses in assessing rater reliability. *Psychological bulletin*. 1979;86(2):420.

32. Li Feng, Qiuting Wen, Chenchan Huang, Angela Tong, Fang Liu, Hersh Chandarana. GRASP-Pro: imProving GRASP DCE-MRI through self-calibrating subspace-modeling and contrast phase automation. *Magnetic resonance in medicine*. 2020;83(1):94-108.

33. Victor Murray, Syed Siddiq, Christopher Crane, Maria El Homsi, Tae-Hyung Kim, Can Wu, Ricardo Otazo. Movienet: Deep space–time-coil reconstruction network without k-space data consistency for fast motion-resolved 4D MRI. *Magnetic resonance in medicine*. 2024;91(2):600-614. doi:10.1002/mrm.29892

34. Melanie Schellenberg, Anthony Mekhanik, Victor Murray, Ricardo Otazo. SELFIE: SElf-supervised Learning for Fast dynamIc golden-anglE radial MRI reconstruction with auto-extracted representations. In: *Proceedings of the International Society for Magnetic Resonance in Medicine Annual Meeting (ISMRM)*. 2025:1188.

35. Elena Vinogradov, Jochen Keupp, Ivan E Dimitrov, Stephen Seiler, Ivan Pedrosa. CEST-MRI for body oncologic imaging: Are we there yet? *NMR in Biomedicine*. 2023;36(6):e4906. doi:10.1002/nbm.4906

36. Ouri Cohen, Victoria Y Yu, Kathryn R Tringale, Robert J Young, Or Perlman, Christian T Farrar, Ricardo Otazo. CEST MR fingerprinting (CEST-MRF) for brain tumor quantification using EPI readout and deep learning reconstruction. *Magnetic Resonance in Medicine*. 2023;89(1):233-249. doi:10.1002/mrm.29448

37. Or Perlman, Christian T Farrar, Hye-Young Heo. MR fingerprinting for semisolid magnetization transfer and chemical exchange saturation transfer quantification. *NMR in Biomedicine*. 2023;36(6):e4710. doi:10.1002/nbm.4710

38. Jorge Zavala Bojorquez, Stéphanie Bricq, Clement Acquitter, François Brunotte, Paul M Walker, Alain Lalande. What are normal relaxation times of tissues at 3 T? *Magnetic resonance imaging*. 2017;35:69-80. doi:10.1016/j.mri.2016.08.021

39. Michael T McMahon, Assaf A Gilad, Jinyuan Zhou, Phillip Z Sun, Jeff WM Bulte, Peter CM van Zijl. Quantifying exchange rates in chemical exchange saturation transfer agents using the saturation time and saturation power dependencies of the magnetization transfer effect on the magnetic resonance imaging signal (QUEST and QUESP): pH calibration for poly-L-lysine and a starburst dendrimer. *Magnetic Resonance in Medicine: An Official Journal of the International Society for Magnetic Resonance in Medicine*. 2006;55(4):836-847. doi:10.1002/mrm.20818

# Supporting Information

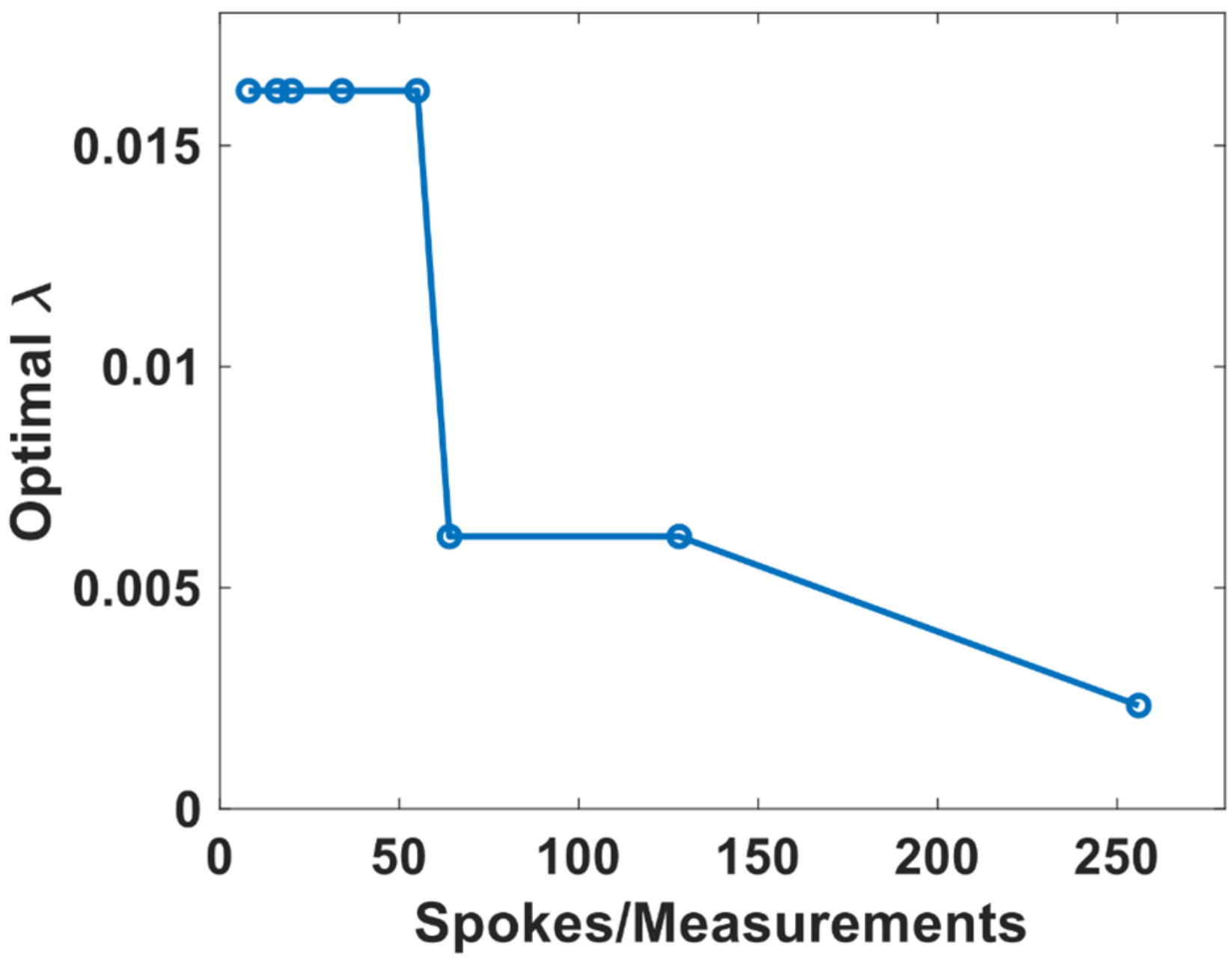

***Figure S1***: *Optimal LRR regularization parameter λ for different values of the number of spokes per measurement.*

**Spoke/Measurement:**

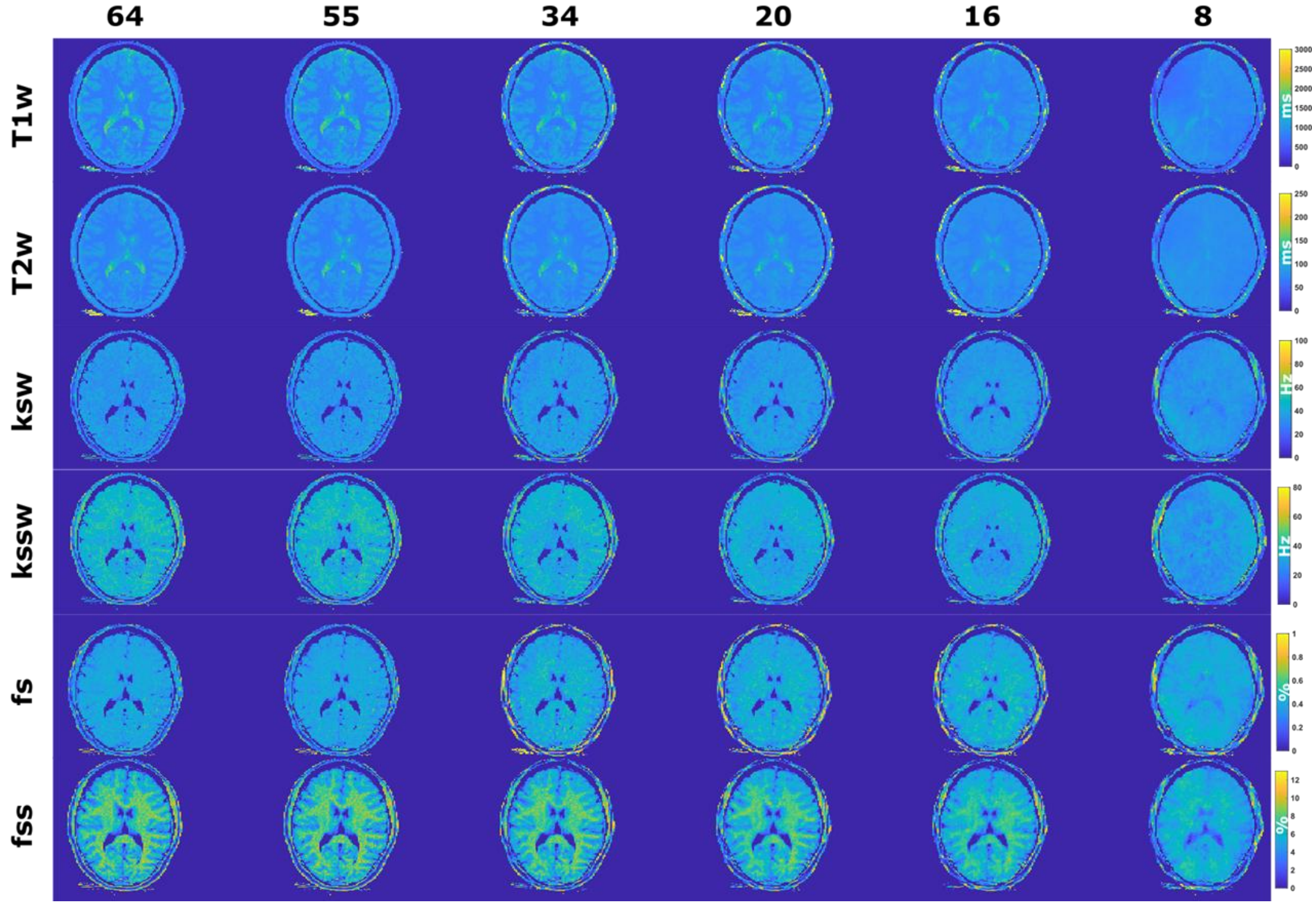

***Figure S2****: Tissue map comparisons for different retrospective undersampling factors in a representative slice shown using the parula colormap to better highlight differences between the various cases. Heavy undersampling (e.g. spokes/measurement=8) resulted in significant blurring, loss of detail, and increased tissue parameter errors. Light to moderate undersampling produced minimal blurring and negligible changes in tissue parameters.*

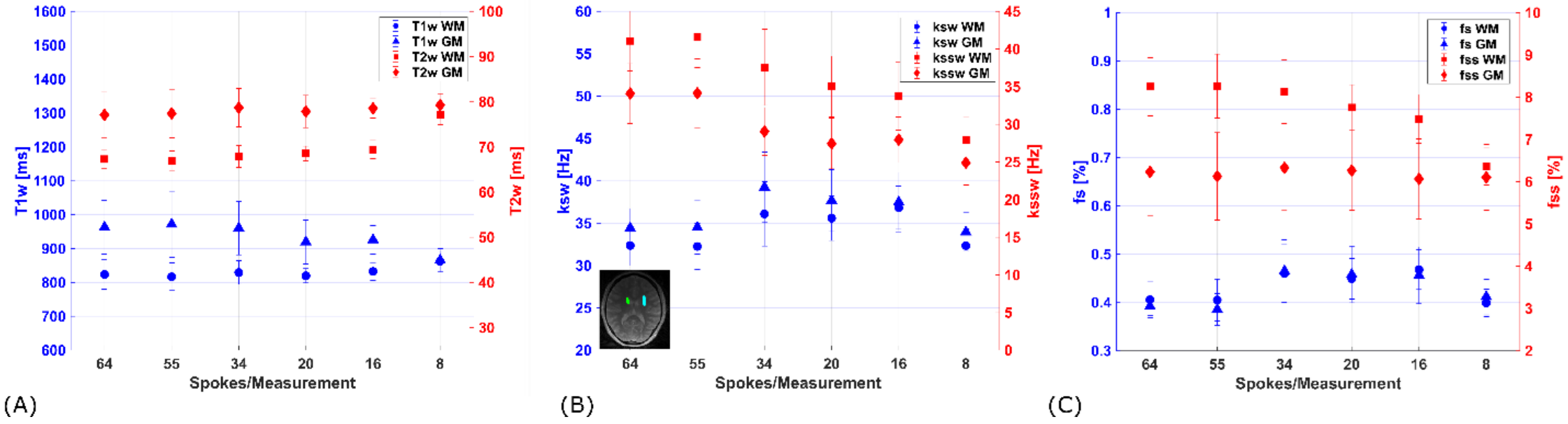

***Figure S3****: Mean ± SD of tissue parameters from retrospectively undersampled acquisitions for gray matter (GM) and white matter (WM) shown on a two-sided plots. Shown are the water T1 (T1w; A), water T2 (T2w; B), amide exchange rate (ksw), semisolid exchange rate (kssw), and volume fraction (fs and fss) values (C). The locations of the WM and GM ROIs are shown inset in panel (B). Fewer spokes reduce scan time but can compromise the accuracy.*

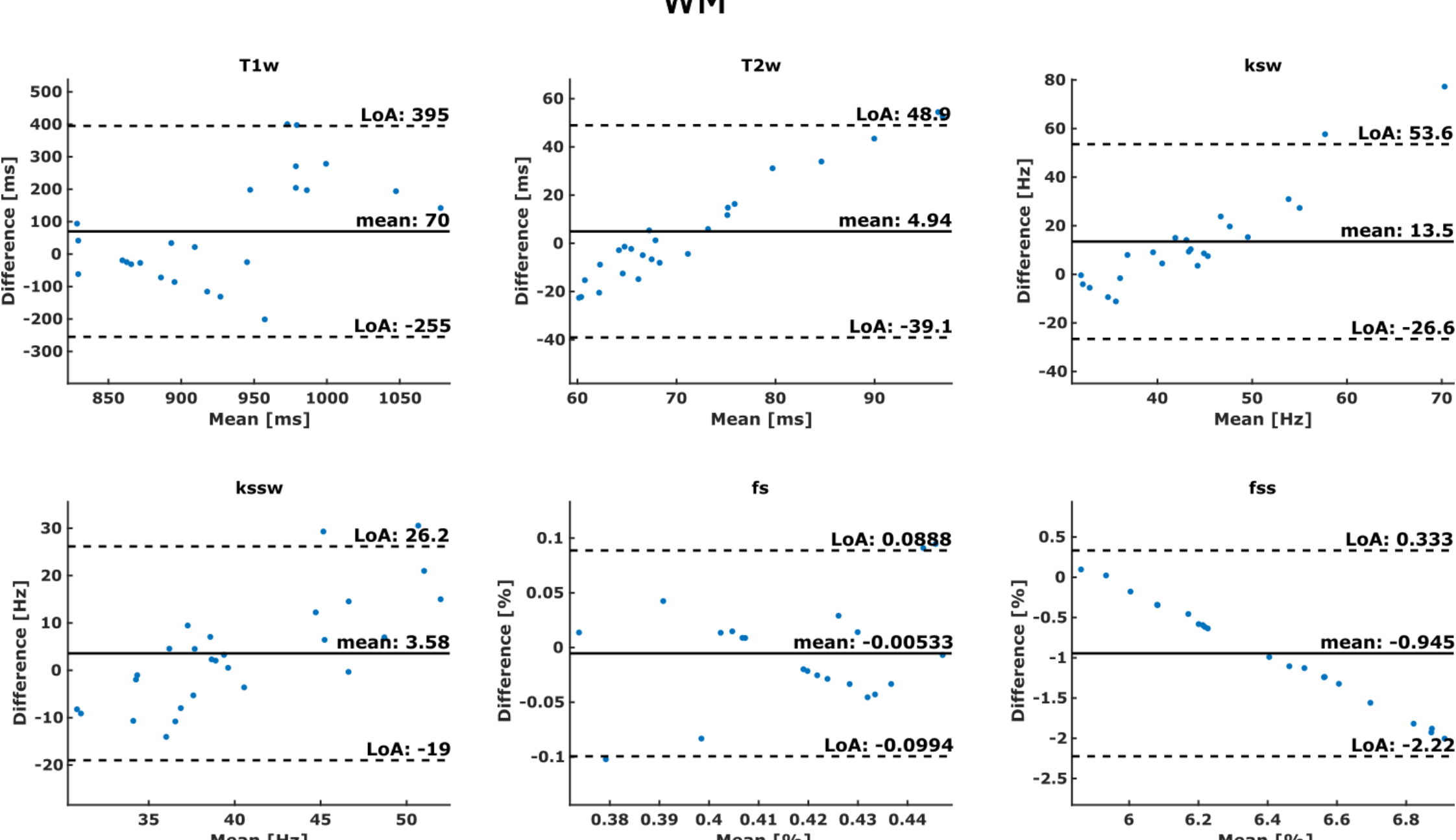

***Figure S4****: Bland–Altman analysis comparing tissue parameter values in WM obtained using the radial CEST-MRF and multi-power fitting. The difference between the two methods is plotted against the mean tissue parameter values for each measurement. The solid horizontal line represents the mean bias, while the dashed lines indicate the 95% limits of agreement (mean bias ± 1.96 SD). The small mean bias and narrow limits of agreement demonstrate good agreement between the two methods for tissue parameter quantification.*

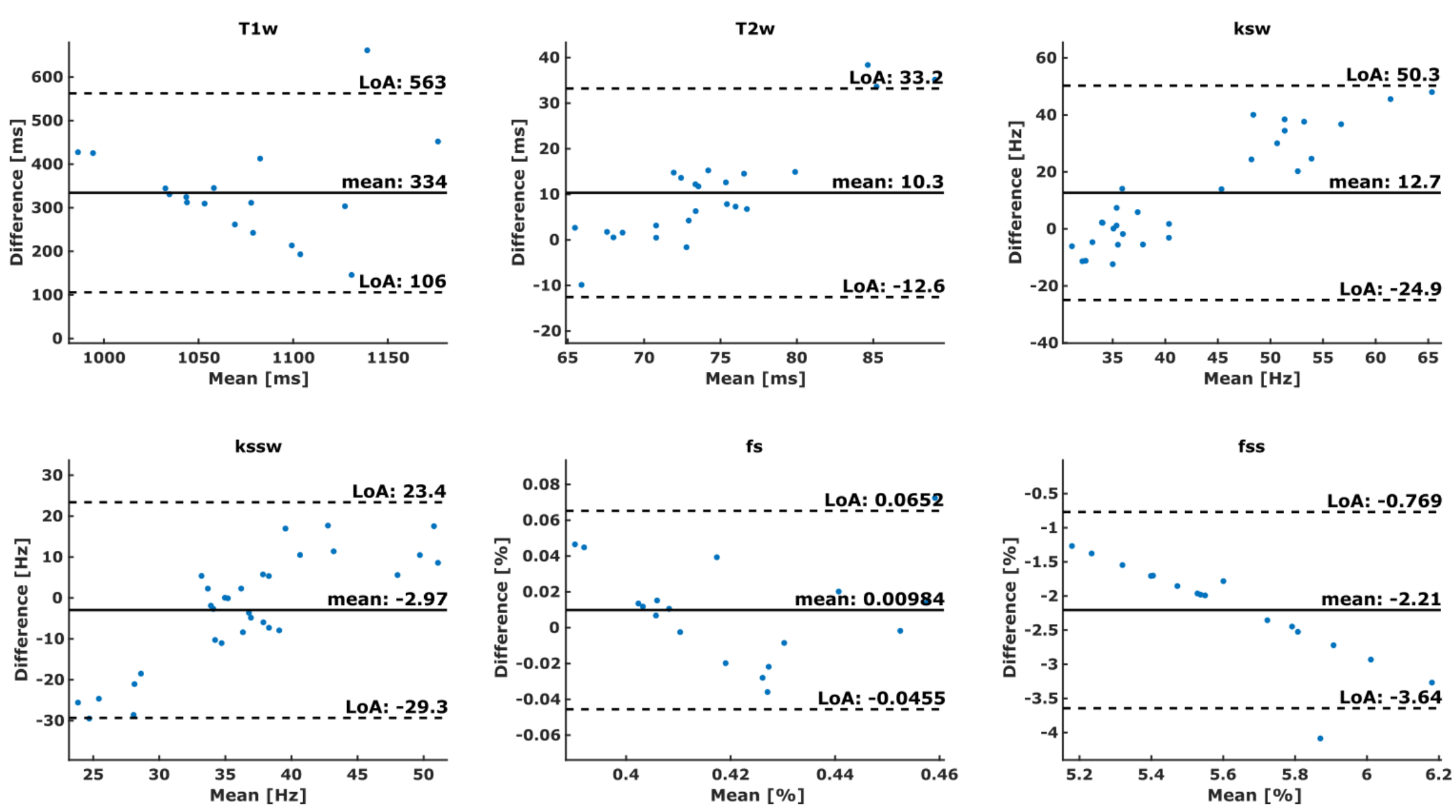

***Figure S5**: Bland–Altman analysis comparing tissue parameter values in GM obtained using the radial CEST-MRF and multi-power fitting. The small mean bias and narrow limits of agreement demonstrate good agreement between the two methods for tissue parameter quantification.*

**Table S1**: Fitting parameters used for the multi-power Z-spectrum CEST NLS curve fitting.

| Parameter | Description | Fitting bounds (lower, upper) |
|---|---|---|
| *Water* | | |
| T1w [ms] | Water longitudinal relaxation time | 400, 3000 |
| T2w [ms] | Water transverse relaxation time | 1, 600 |
| *CEST* | | |
| T1s [ms] | Semisolid longitudinal relaxation time | 1450 |
| T2s [ms] | Semisolid transverse relaxation time | 1 |
| ksw [Hz] | Amide exchange rate | 1, 100 |
| fs [%] | Amide volume fraction | 0, 0.91 |
| *MT* | | |
| T1ss [ms] | Semisolid longitudinal relaxation time | 1450 |
| T2ss [ms] | Semisolid transverse relaxation time | 1, 10 |
| kssw [Hz] | Semisolid exchange rate | 1, 100 |
| fss [%] | Semisolid volume fraction | 0, 13.64 |
| *NOE* | | |

| T1a [ms] | Amide longitudinal relaxation time | 1450 |
|---|---|---|
| T2a [ms] | Amide transverse relaxation time | 1, 100 |
| ka [Hz] | NOE exchange rate | 1, 50 |
| fa [%] | NOE volume fraction | 0, 0.55 |
| *Instrumental* | | |
| B1 [a.u.] | Transmit field inhomogeneity | 0.4, 1.2 |
| B0 [ppm] | Static field inhomogeneity | -0.5, 0.5 |